\input harvmac
\input epsf
\def\Title#1#2{\rightline{#1}\ifx\answ\bigans\nopagenumbers\pageno0\vskip1in
\else\pageno1\vskip.8in\fi \centerline{\titlefont #2}\vskip .5in}

%
%
\ifx\epsfbox\UnDeFiNeD\message{(NO epsf.tex, FIGURES WILL BE IGNORED)}
\def\figin#1{\vskip2in}
\else\message{(FIGURES WILL BE INCLUDED)}\def\figin#1{#1}
\fi
\def\Fig#1{Fig.~\the\figno\xdef#1{Fig.~\the\figno}\global\advance\figno
 by1}
%
%
%
%
\def\ifig#1#2#3#4{
\goodbreak\midinsert
\figin{\centerline{\epsfysize=#4truein\epsfbox{#3}}}
\narrower\narrower\noindent{\footnotefont
{\bf #1:}  #2\par}
\endinsert
}

%
%
\font\ticp=cmcsc10

\def\ajou#1&#2(#3){\ \sl#1\bf#2\rm(19#3)}
\def\jou#1&#2(#3){,\ \sl#1\bf#2\rm(19#3)}
\def\hf{{1\over 2}}

\def\frac#1#2{{#1\over#2}}

\def\cald{{\cal D}}

\def\calo{{\cal O}}

\def\caln{{\cal N}}
\def\calO{{\cal O}}

\def\im{{\rm Im}}
\def\vecm{{\vec m}}
\def\vecx{{\vec x}}
\def\nlm{{nl\vecm}}
\def\vac{\vert0\rangle }
\def\ehat{{\hat e}}

%
\lref\SuTo{L. Susskind and N. Toumbas, ``Wilson Loops As Precursors,''
hep-th/9909013, {\sl Phys.Rev. } {\bf D61} (2000) 044001.}  
\lref\SBGBH{See, {\it e.g.} 
S.B. Giddings, ``The black hole information paradox,''
hep-th/9508151\semi ``Quantum Mechanics of Black Holes,''
hep-th/9412138\semi A. Strominger, ``Les Houches Lectures on Black Holes,''
hep-th/9501071, NATO Advanced Study Institute: Les Houches Summer School, 
Session 62: {\sl Fluctuating Geometries in Statistical 
Mechanics and Field Theory}, F. David, P. Ginsparg, and J. Zinn-Justin
(eds.).}
\lref\tHoo{G. 't Hooft, ``Dimensional Reduction in Quantum Gravity,''
gr-qc/9310026.}
\lref\Suss{L. Susskind, ``The World as a Hologram,'' hep-th/9409089,
{\sl J. Math. Phys.} {\bf 36} (1995) 6377.}
\lref\Malda{J. Maldacena, ``The Large N Limit of Superconformal Field
Theories and Supergravity,'' hep-th/9711200, {\sl
Adv. Theor. Math. Phys.} {\bf 2} (1998) 231.}
\lref\PST{J. Polchinski, L. Susskind, N. Toumbas, ``Negative Energy,
Superluminosity and Holography,'' hep-th/9903228 {\sl Phys. Rev.} {\bf
D60} (1999) 084006.}
\lref\SoTo{L. Susskind, and N. Toumbas, ``Wilson Loops as Precursors,''
hep-th/990901, {\sl Phys. Rev.} {\bf D61} (2000) 044001.}
\lref\LMR{J. Louko, D. Marolf, and S. F. Ross, ``On geodesic propagators and black hole 
holography,'' hep-th/0002111, {\sl Phys. Rev.} {\bf D62} (2000) 044041.} 
\lref\Polc{J. Polchinski, ``S-Matrices from AdS Spacetime,''
hep-th/9901076.}
\lref\Sussk{L. Susskind, ``Holography in the Flat Space Limit,''
hep-th/9901079.}
\lref\FSS{S.B. Giddings, ``Flat-space scattering and bulk locality in
the  AdS/CFT correspondence,'' hep-th/9907129, {\sl Phys. Rev.}
{\bf D61} (2000) 106008.}
\lref\Azam{M. Azam, ``Local gauge-invariant generators for Wilson
loops,''\ajou Phys. Rev. &D40 (89) 3541.}
\lref\BGL{V. Balasubramanian , S. B. Giddings, and  A. Lawrence,
``What Do  CFTs Tell Us About Anti-de Sitter Spacetimes?''
hep-th/9902052 {\sl JHEP} {\bf 9903} (1999) 001.}
\lref\BSM{S. B. Giddings, ``The boundary S-matrix and the AdS to CFT
dictionary,'' hep-th/9903048, {\sl Phys. Rev. Lett.} {\bf 83} (1999) 2707.}
\lref\BCFM{D. Berenstein, R. Corrado, W. Fischler, and  J. Maldacena, ``The
Operator Product Expansion for Wilson Loops and Surfaces in the Large
N Limit,'' hep-th/9809188, {\sl Phys. Rev.} {\bf D59} (1999) 105023.}
\lref\Mald{J. Maldacena, ``Wilson loops in large N field theories,''
hep-th/9803002 {\sl Phys. Rev. Lett.} {\bf 80} (1998) 4859.}
\lref\BDHM{T. Banks, M. R. Douglas, G. T. Horowitz, and E. Martinec, ``AdS
Dynamics from Conformal Field Theory,'' hep-th/9808016.}
\lref\HoPo{G. T. Horowitz and J. Polchinski, ``A Correspondence Principle for Black Holes
and Strings,'' hep-th/9612146, {\sl Phys. Rev.} {\bf D55} (1997) 6189\semi
``Self Gravitating
Fundamental Strings,'' hep-th/9707170, {\sl Phys. Rev.} {\bf D57}
(1998) 2557.}
\lref\GrMe{D.J. Gross and P.F. Mende, ``The
high-energy behavior of string scattering amplitudes,'' {\sl
Phys. Lett.} {\bf 197B} (1987) 129\semi ``String theory beyond the
Planck Scale,'' {\sl Nucl. Phys} {\bf B303} (1988) 407.}
\lref\Hawk{S.W. Hawking, ``Particle Creation by Black Holes,'' {\sl
Commun. Math. Phys.} {\bf 43} (1975) 199.}
\lref\GiNe{S. B. Giddings and W. M. Nelson, ``Quantum Emission from
Two-Dimensional Black Holes,'' hep-th/9204072, {\sl Phys. Rev.} {\bf
D46} (1992) 2486.}
\lref\Yone{T. Yoneya, ``Space-Time Uncertainty and Noncommutativity in
String Theory,'' hep-th/0010172, and references therein.}
\lref\LPSTU{D. A. Lowe, J. Polchinski, L. Susskind, L. Thorlacius, and
J. Uglum, ``Black Hole Complementarity vs. Locality,'' hep-th/9506138,
{\sl Phys. Rev.} {\bf D52} (1995) 6997.}
\lref\LaMa{A. Lawrence and E. Martinec, ``Black Hole Evaporation along
Macroscopic Strings,'' hep-th/9312127, {\sl Phys. Rev.} {\bf D50} (1994) 2680.}
\Title{\vbox{\baselineskip12pt
\hbox{hep-th/0103231}\hbox{NSF-ITP-01-23}
}}
{\vbox{\centerline{Precursors, black holes,}\vbox{\centerline {and
a locality bound}}
}}
\centerline{{\ticp Steven B. Giddings}$^{ab}$\footnote{$^{\dagger}$}
{giddings@physics.ucsb.edu} and {\ticp Matthew 
Lippert}$^a$\footnote{$^{\ast}$}
{lippert@physics.ucsb.edu}}
\bigskip\centerline{${}^a$ {\sl Department of Physics}}
\centerline{\sl University of California}
\centerline{\sl Santa Barbara, CA 93106-9530}
\bigskip\centerline{${}^b$ {\sl Institute of Theoretical Physics}}
\centerline{\sl University of California}
\centerline{\sl Santa Barbara, CA 93106-4030}

\bigskip
\centerline{\bf Abstract}
We revisit the problem of precursors in the AdS/CFT correspondence.
Identification of the precursors is expected to improve our understanding
of the tension between holography and bulk locality and of the resolution
of the black hole information paradox.  Previous arguments that the
precursors are large, undecorated Wilson loops are found to be flawed.   We argue
that the role of precursors should become evident when one saturates a
certain locality bound.  The spacetime uncertainty principle is a direct
consequence of this bound.

\Date{}

\newsec{Introduction}

The puzzles of quantum gravity become sharply focussed with the black hole
information paradox\refs{\SBGBH}, which arises when considering the
fate of quantum mechanical information which falls into a black hole.
Destruction of the information would sacrifice quantum mechanics and
would apparently lead to physics that violates energy conservation,
while escape of the information in Hawking radiation would appear to
violate locality.

This difficult situation led to the postulated {\it holographic
principle}\refs{\tHoo,\Suss}, which holds that in a real sense the
information can be thought of as stored in degrees of freedom at the
surface of  the black hole.  This principle conflicts with
locality as usually formulated in quantum field theory, but only in extreme
circumstances; at long distances and low energies the world should
remain effectively local.

The holographic principle has found a concrete realization in Maldacena's
proposed AdS/CFT correspondence\refs{\Malda}, which asserts that
string theory in the whole of AdS spacetime has an equivalent description
as dynamics of a large-N super Yang-Mills theory on the boundary of 
that spacetime.

If true, this equivalence says that all information inside AdS can be
equivalently described by a state of the boundary.  This would include
information that from the bulk perspective has not had time to causally
reach the boundary.  An example would be a bomb detonated at the center of
AdS; from the bulk perspective the information  from the bomb
should not reach the
boundary until a time comparable to the AdS radius $R$, but equivalence
with the boundary theory implies that this information should be somehow
encoded in the boundary state the moment the bomb goes off.  Polchinski,
Susskind, and Toumbas\refs{\PST} formulated the important question of
identifying these boundary variables in which the information is encoded
and coined the name {\it precursors} to describe them.

Going one step further, if observation of precursors allows one to measure
information that should be causally inaccessible from the bulk perspective,
precursors should allow one to measure information
inside a black hole in anti-de Sitter space.  Indeed, according to the
holographic principle, black hole formation and evaporation is a unitary
process and, by AdS/CFT, should be fully encoded at all times in the
boundary CFT.  For this reason it would be extremely interesting to
identify the precursor fields and use them to chart the internal dynamics
of a black hole.

Susskind and Toumbas\refs{\SuTo} have made the concrete proposal that the
precursor fields are large Wilson loops and have presented calculations
purporting to show that these Wilson loops indeed allow boundary measurements
that would na\"\i vely be forbidden by bulk locality.  In particular, in
the case of the explosion mentioned above, measurement of a Wilson loop of
size $a$ would allow a detection of the explosion at a time of order $a$ before
the light cone of the explosion reaches the boundary of AdS.  

It should be noted that it is debatable to what extent such an observation
-- even if possible -- constitutes observing the explosion outside its
light cone.  To foresee the explosion by a time $a$ requires a Wilson loop
of size $a$, and it would appear to take a time $a$ to actually know that
the Wilson loop has been measured -- the data from the detectors along the
loop would have to be sent to some central location for comparison.\foot{We
thank J. Preskill for discussions on this point.}   However, as we'll
discuss, one could also imagine using Wilson loops to measure events inside
a black hole.  In this case, any measurement would be extremely
interesting, since the time it would take the information to escape
classically is infinite.

In this paper we investigate these claims more closely. There is a purely
field theoretical calculation analogous to that of \SuTo\ that also
seems to indicate that observation of a bilinear of local operators
allows one to likewise measure the explosion acausally.  However, we
know from field theory causality that this cannot be correct.  We
trace the conflict to an incorrect identification of the saddlepoint
in an integral in an analysis analogous to \SuTo.  A closer inspection
of the string theory expression of \SuTo\ shows that the
saddlepoint has been incorrectly identified there as well, invalidating
that analysis.\foot{A related discussion has appeared in \refs{\LMR}.} 
While in field theory we know that the exact calculation
predicts that an event cannot be measured outside its light cone, we do not
yet know how to do an analogous calculation in string theory without
computing off-shell quantities.  We outline a possible calculation 
and comment on our expectations for the result and its connection to black
holes.  Our results raise serious questions about the identification of
large Wilson loops as prescursors.

We therefore return to consider the motivations for holography and its
attendant breakdown of locality.  An underlying principle is that locality
should fail when we attempt to make measurements in which black holes or
strings are created.  We propose a concrete criterion for such a 
locality bound and outline its possible implications for the
problem of precursors in AdS/CFT.  We also discuss the connection to the
problem of holographically encoding the internal state of a black hole.

We close our introduction by mentioning another logical, though
heretical possibility.  It may be that the AdS/CFT correspondence is
not a 1-1 map; it could be that the CFT does not include all the
information encoding bulk physics, for example on scales less than the
AdS radius scale, $R$.  Indeed, attempts \refs{\Polc,\Sussk,\FSS} 
to extract such information from correlators in the CFT have run into
difficulties.  If there are such missing degrees of freedom they might be
related to the precursors.

The outline of this paper is as follows.  Section two gives a more detailed
review of the problem of identifying the precursors.  Section three
investigates the proposal of \refs{\SuTo} that the precursors are large
Wilson loops, finds a flaw in that analysis, and proposes a refined
calculation that would be necessary to demonstrate the validity of that
proposal.  This section can be skipped by those who don't believe that
large Wilson loops are precursors. 
Section four makes the alternative proposal that the precursors
are related to observations at sufficiently high energies for locality to
break down.  We give a concrete suggestion for a criterion for such
a locality bound.  Section five discusses the relation of both
proposals to the problem of charting the internal dynamics of a black hole,
and in section six we give comments and conclusions.

\newsec{The problem of precursors}

We begin by describing the problem of precursors in some more detail, in
the process reviewing some of the basics of AdS/CFT.  We begin with 5d
anti-de Sitter space in global coordinates, 
\eqn\adsmet{ds^2= {R^2\over \cos^2\rho}(-d\tau^2 + d\rho^2 + \sin^2\rho \,
d\Omega_3^2)\ .}
We denote the bulk coordinates as $x=(\tau,\rho,\Omega)$ and the boundary
coordinates as $b=(\tau,\Omega)$.  Now, imagine there is a source
of one of the fields, say the dilaton $\phi$, at the center of AdS at
time $\tau=0$.   For concreteness, we idealize this source as point-like
in space and time,
\eqn\source{j(x) = j \delta(x)\ .}
In the bulk language, this creates a state $\vert j\rangle_B$.  We will work in
the field theory approximation and use an interaction picture with $j\phi$ treated
as the interaction; the state is then 
\eqn\bulksource{\vert j, t\rangle_B = \exp\left\{i\int^t dV_x \, j(x) \phi(x)\right\}
\vert0\rangle_B} 
where $\vert0\rangle_B$ is the bulk vacuum.  For $t>0$, \source\ and \bulksource\
give
\eqn\bulksourcespecific{\vert j\rangle_B = \exp\left\{i j \phi(0) \right\}
\vert0\rangle_B.}

Fields in AdS map to operators on the boundary through the map  
\eqn\opfield{{\cal O}_\Delta(b) \leftrightarrow \lim_{\rho\rightarrow\pi/2}
(\cos\rho)^{-\Delta} \phi_\Delta(x) }
where $x\rightarrow b$ is the limiting point on the boundary and
$\Delta$ represents the CFT dimension of the 
field.  This correspondence induces a map from bulk to
boundary states that we spell out further in section 4.  In particular,
corresponding to \bulksource, the boundary
state for times $t>0$ takes the general form
\eqn\bdystate{\vert j\rangle_\partial = \exp\left\{ i\int db f(b)
\calO(b)\right\} \vert0\rangle_\partial}
where $\vac_\partial$  is the boundary vacuum, $\calO$ is the operator
$\frac{1}{N}{\rm Tr} \, F^2$ corresponding to the dilaton, and $f(b)$ is a function
determined by $j(x)$.

In the context of quantum field theory in the bulk, 
it is clear that no information about the source reaches the boundary until time
$\tau=\pi/2$, when the light cone of the source meets the
boundary. On the other hand, since according to the holographic proposal 
the boundary theory contains all the
information of the bulk theory, 
\bdystate\ should contain the information about the source before this
time.  For example, instead of \source, we might imagine the source sending
a message encoded in variations of $j(x)$ over a short time around $\tau=0$,
and the boundary state should contain all the information of this source.
Simply put, the question of identifying the precursors is the question of
understanding what degrees of freedom and observables in the boundary
theory encode this information.  Answering this question is an important
step towards decoding the hologram and, in particular, towards understanding
how approximate bulk locality is encoded and ultimately fails.

\ifig{\Fig\Shell}{Identification of local precursor fields in AdS/CFT may
allow measurements outside the light cone of a source, violating na\"\i ve
bulk locality.}{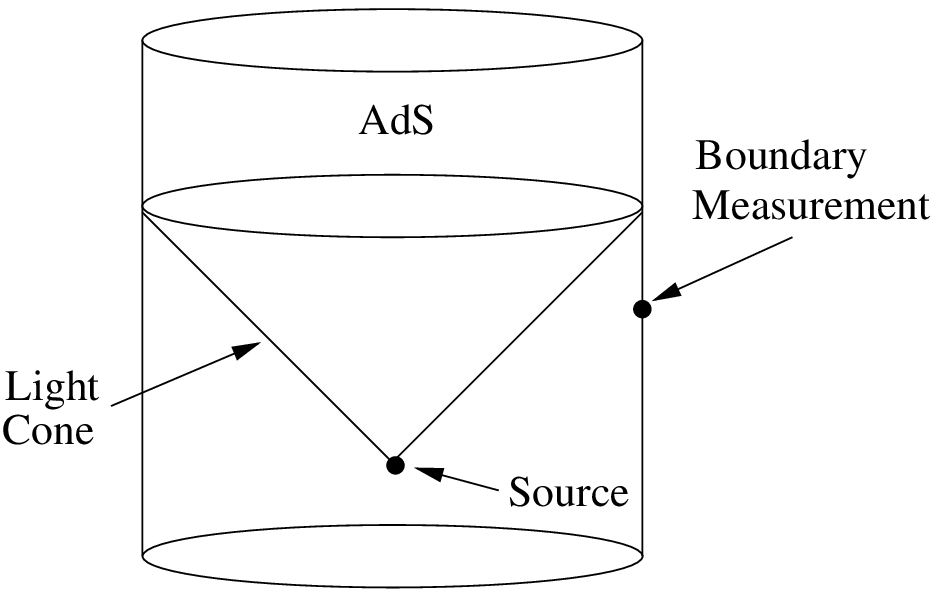}{2.0}

Since the boundary theory is $\caln=4$ super Yang-Mills, we know that a
basis for all observables is given by the set of all Wilson loops.
Equivalently, each Wilson loop can be expanded (at least formally) in terms
of an infinite series of local operators at a point\refs{\Azam}.  The
question, therefore, is to identify which of these Wilson loops or local
operators one should measure to detect information outside the light cone of
the source.

In particular, consider more closely the bulk/boundary correspondence for
Wilson loops.  We know that correlators of local boundary operators map to the
AdS analog of the S-matrix\refs{\BGL,\BSM} (called the {\it boundary 
S-matrix} in \BSM) and would like a corresponding statement for Wilson loops.  
We expect that this map between
correlators and S-matrices also extends to a statement for Wilson loops,
namely that a correlator of Wilson loops in the boundary
theory corresponds to a boundary S-matrix for large loops of string.
Note that, of course, at least at the formal level, an arbitrary Wilson loop
can be decomposed into an infinite sum of local, but arbitrarily
high-dimension operators at a
point\refs{\Azam}, which we expect to correspond to representing a large
string in terms of its modes.

Although we know of no complete and usable string field theory
description of AdS space, we will find it useful to explain our
picture in string field theory terms.
At least perturbatively, the ultimate expressions we will consider can
then be rewritten as first-quantized integrals over the resulting string
world sheets. 

The string field $\Phi[x(\sigma)]$ is a functional of string loops
$x(\sigma)$, as well as ghosts and other fields which we suppress.  
Extending the Ansatz of \refs{\Mald}, a Wilson loop operator in
the boundary theory is identified, in analogy to \opfield, as the boundary
limit of the string field operator, which creates a string loop:
\eqn\opstring{W(C) \leftrightarrow
\lim_{x(\sigma)\rightarrow C} Z[x(\sigma)] \Phi[x(\sigma)]}
where $Z[x(\sigma)]$ is a (infinite) normalization factor analogous to that needed
for point-like operators.
Furthermore, note
that the dilaton field operator $\phi$ is a projection of this string field
to the dilaton mode.

It was proposed in \SuTo\ that {\it large Wilson loops} act as the
precursors:  In order to measure the source a time $a$ before its light cone
reaches the boundary, one should measure the expectation value of a spatial
boundary
Wilson loop $W(C)$ with size of order $a$,
\eqn\exptvl{ {}_\partial\langle j\vert W(C) \vert j\rangle _\partial\ .}
In the next section we will examine this proposal more closely and find a
flaw in the analysis of \SuTo, reopening the question of finding the
precursors.

\newsec{Large Wilson loops as precursors?}

\subsec{Review and reformulation}

Considering the source \source\ of the preceding section, the authors of \SuTo\
advocate that we consider making an
observation using a large Wilson loop $W(C)$,
\eqn\bdyobserv{{}_\partial\langle j\vert W(C)\vert j\rangle_\partial\ ,}
where the curve $C$ lies completely outside the light cone of the source.
This correlator can be calculated to linear order in $j$ by expanding
\bdystate\ and compared with the vacuum expectation value for the Wilson
loop.  Non-vanishing of the resulting difference,
\eqn\acausal{i\int db f(b) \, {}_\partial\langle 0\vert [W(C), \calO(b)] 
\vert 0\rangle_\partial }
would be an indicator that information had been measured outside the light
cone of the source.

Ref.~\SuTo\ infers general properties of $f$, and uses an (approximate) 
calculation of
\eqn\correlator{
{}_\partial\langle 0\vert [W(C), \calO(b)]
\vert 0\rangle_\partial}
given by Berenstein, Corrado, Fischler, and Maldacena\refs{\BCFM}.  
Combining these answers yields a non-vanishing answer for \acausal,
purporting to demonstrate that the Wilson loop $W(C)$ is indeed
capable of measuring the boundary effects of the source outside its light cone.

This approach proceeds via
a calculation in the boundary field theory, though the
boundary source function $f$ is inferred from the bulk source $j$ and the
boundary correlator \correlator\ is inferred in \refs{\BCFM} from a bulk
computation.  It is equivalent, and more straightforward, to perform all
calculations directly in the bulk theory, as we will now do.

Again, working to linear order in the source $j$, the bulk analogue 
to \acausal\ is
\eqn\bcausal{\eqalign{ i\lim_{x(\sigma)\rightarrow C} Z[x(\sigma)]&    
\int dx \, j(x) \, {}_B\langle 0\vert [\Phi[x(\sigma)],\phi(x)] \vert 0\rangle_B
\cr = & \, ij \lim_{x(\sigma)\rightarrow C} Z[x(\sigma)] \, {}_B\langle 0\vert
[\Phi[x(\sigma)], \phi(0)]\vert 0\rangle_B\ .} }
By hermiticity of the operators, we can then rewrite the expectation value
of the commutator as 
\eqn\Impart{ {}_B\langle 0\vert [\Phi[x(\sigma)],\phi(0)] \vert 0\rangle_B
= 2i \, \im  \, {}_B\langle 0\vert \Phi[x(\sigma)] \phi(0) \vert 0\rangle_B\ .}
We need to compute the string two-point
function from the point-like dilaton state at $x=0$ to the boundary loop
$x(\sigma)\rightarrow C$.  Such far off-shell calculations in string theory
are notoriously difficult.  However, the analysis of \Mald\ and \BCFM\
suggests that the answer is well approximated by a saddlepoint.  Indeed, it
would seem that the obvious extremal surface corresponding to this
configuration is a minimal area surface spanning the loop, which for
convenience we take to be purely spacelike, and then a thin tube -- or
dilaton propagator -- connecting the origin to a point on this surface.  

\ifig{\Fig\LineVsTube}{The string world sheet  can be approximated by a dilaton propagator attached to a minimal
surface.}{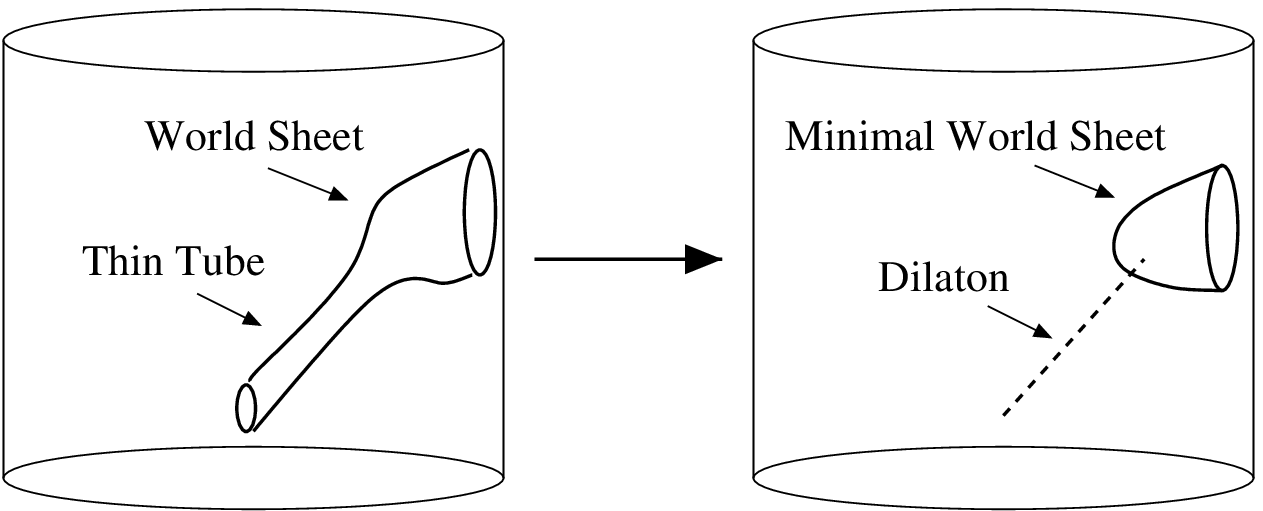}{2.0}

Indeed, specifically considering a circular spacelike Wilson loop and 
directly following \BCFM\ ({\it c.f.} eq. 4.9), 
with the minor modification that the point-like operator sits in the bulk,
we are lead to an expression
\eqn\wrongsadd{ \lim_{x(\sigma)\rightarrow C} Z[x(\sigma)] \, {}_B\langle0\vert 
\Phi[x(\sigma)]
\phi(0) \vert 0\rangle_B \propto \int d {\cal A'} K_B(0,x')\ ,}
where the integral is over points $x'$ on 
the minimal surface spanning the loop and $K_B$
is the bulk AdS propagator.  

By standard field theory causality in AdS space, the bulk propagator is
purely real outside the light cone but has an imaginary piece inside the
light cone.  As in \SuTo, large enough Wilson loops on the boundary,
but outside the light cone, will produce spanning surfaces that enter the
interior of the light cone.  This leads to a non-vanishing imaginary part
of \wrongsadd\ and hence the appearance that the Wilson loop is sensitive to
information not accessible by usual causal observations.  If one wants to
measure the source at time $a$ before its light cone reaches the boundary, a
rough criterion for the relevant Wilson loops is that they should have
radius $\sim a$; this condition allows the spanning minimal surface to dip
into the interior of the light cone.

\subsec{A field theory model}

We now discuss a pure field theory analog of the Wilson loop analysis of
the previous subsection.  Suppose that instead of a Wilson loop, the
boundary observer measures a bilocal operator $\calO(b)\calO(b')$.  Let us
consider a simple toy model  of a field theory with a massless scalar
$\phi$ coupled to a scalar $\psi$ of mass $M$ through a purely cubic
interaction, $g\int dx \, \phi(x)\psi^2(x)$.  Consider a source at $x=0$
as in eq.~\source, and suppose that the boundary points $b$ and $b'$
are spacelike separated; for concreteness take them to be at equal
global AdS times, and furthermore assume that they are both outside
the light cone of the source. 

With the obvious substitutions in the above steps, the result for the
observation of the bilinear takes the form
\eqn\ftobs{\eqalign{ i\int db'' f(b'') & \, {}_\partial\langle0\vert 
[\calO_\psi(b) \calO_\psi(b'), \calO_\phi(b'')] \vert0\rangle_\partial\cr &=
-2j\lim_{x\rightarrow b}\lim_{x'\rightarrow b'} (\cos\rho)^{-\Delta}
(\cos\rho')^{-\Delta} \, \im \, {}_B \langle0\vert \psi(x)\psi(x')\phi(0)
\vert0\rangle_B\ , }}
analogous to \acausal.

Following steps identical to those of \refs{\SuTo, \BCFM}, we approximate
the expression \ftobs\ as follows.  
At tree level in the interaction parameter $g$, it contains
\eqn\feynexp{{\rm Im} \,
{}_B\langle0\vert \psi(x)\psi(x') \phi(0) \vert0\rangle_B =-\im\, ig
\int dV_y \, K_B(y,x;M) K_B(y,x';M) K_B(0,y;0)\  }
where we have explicitly indicated the mass in the propagator.  We can represent this expression, in analogy to the sum over world sheets,
as a first-quantized functional integral over world lines as shown in
fig.~3.  

\ifig{\Fig\ThreePoint}{The measurement of the source by a bilocal operator
at the boundary can be written in terms of the imaginary part of a
three-point function.}{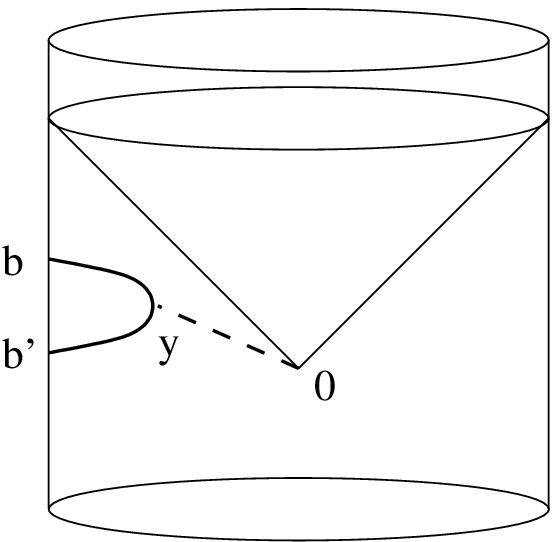}{2.0}

For $M|x-x'|\gg 1$, we expect, completely in analogy with 
\refs{\Mald,\BCFM}, that this is dominated by a configuration with a
minimal line connecting points $x\rightarrow b$ and $x'\rightarrow b'$
and with the $\phi$ propagator
connecting the origin to an arbitrary point along this line.  So, we expect
that
\eqn\wrongft{ {\rm Im} \,
{}_B\langle0\vert \psi(x)\psi(x') \phi(0) \vert0\rangle_B \propto \int dl_y
K_B(0,y)\ ,}
where $y$ is integrated along the minimal curve connecting $x$ to $x'$, in precise
analogy with \wrongsadd.  For large enough separation of $b$ and $b'$, this
minimal curve enters the future light cone of the source, where the bulk
propagator is complex, and thus \wrongft\ picks up a non-vanishing
imaginary part.  The bilinear thus can make measurements outside the light
cone.

The preceding is, of course, utter nonsense.  In intermediate steps,
\ftobs\ was derived from an expression of the form
\eqn\impart{ {}_B\langle0\vert [\psi(x) \psi(x'), \phi(0)]
\vert0\rangle_B = 2i \, {\rm Im} 
{}_B\langle0\vert \psi(x)\psi(x') \phi(0) \vert0\rangle_B\ .}
Since $x$ and $x'$ are spacelike to $0$, the commutator must vanish by
standard field theory causality.  We'll derive an analogous statement in
terms of flat-space Feynman diagrams in the next subsection.

\subsec{Searching for a pass through the mountains}

What went wrong with the approximation analogous to \BCFM, and is the
same problem encountered in the string case?  To answer this, consider
redoing the field theory analysis in 4d Minkowski space;  indeed,
completely analogous reasoning there leads to the conclusion that the
bilinear $\psi(x)\psi(x')$ can measure information for $x$ and $x'$ outside
the light cone of a source at the origin as long as the straight line
connecting them intersects the interior of the light cone to produce a
non-vanishing imaginary part.

To leading order in $g$, the exact expression that we should consider is
\eqn\exactgf{\im \, i \int d^4y \, D(0,y;0) D(y,x;M) D(y,x';M) }
where $D(x,y;m)$ denotes the Feynman propagator of mass $m$.  Although \exactgf\ vanishes by causality as in \impart, we can show it
directly as follows.  In four spacetime dimensions, the Feynman propagators can be
written as
\eqn\propschwinger{
D(x,y;M) = -\frac{i}{8\pi^2} \int_0^\infty dt \, t^{-2} 
e^{{i(x-y)^2}/{2t} - {iM^2t}/{2} - \epsilon t} 
.}
Using this representation of the propagator, the Minkowski version of
the amplitude \feynexp\ becomes
\eqn\unrotated{\eqalign{
&\im \, {}_B\langle0\vert \psi(x)\psi(x') \phi(0) \vert0\rangle_B \cr
&=  \im \, \frac{g}{512\pi^6} \int_0^\infty ds\, dt\, du\, (stu)^{-2}
e^{-iM^2(s+t)/2-\epsilon(s+t+u)} 
\int d^4y \, e^{\frac{i}{2} \left[{(x-y)^2}/{s} + {(x'-y)^2}/{t} + 
{y^2}/{u}\right]}
.}}
We can now perform the Gaussian integral over $y$ exactly.  Because the
resulting expression depends only on $x^2$, $x'^2$, and $(x-x')^2$,
all of which are positive, the integrals over $s$, $t$, and $u$ can be
rotated $s, t, u \to -is, -it, -iu$
without encountering poles.  We now have
\eqn\rotated{\eqalign{
-\im \, \frac{g}{128\pi^4} \int_0^\infty& ds\, dt\, du\, (st+tu+us)^{-2}\cr 
&\exp\left\{-\hf\left[M^2(s+t) + \frac{tx^2 + sx'^2 +
u(x-x')^2}{st+tu+us}\right]\right\} \ 
.}}
The integrals are manifestly real and convergent, so \rotated\  has
vanishing imaginary part and \impart\ is zero to first order in $g$.

\ifig{\Fig\FeynmanDiagram}{The amplitude in \exactgf\ can be represented as
a tree-level Feynman diagram}{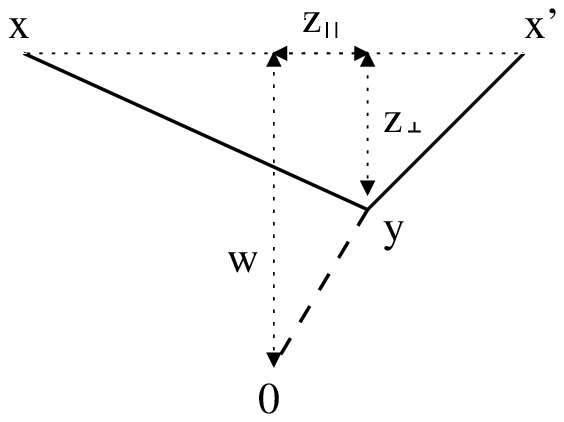}{2.0}

Now, consider instead approximating \exactgf\ in the manner of \BCFM.
For $M|x-x'|\gg1$, the obvious guess is that the integral is dominated
by  $y$ near the line $xx'$.  We work in the approximation  $M|x-y|\gg1$,
$M|x'-y|\gg1$.  Outside the light cone, the massive Feynman propagators
(in four spacetime dimensions) are
\eqn\feynprop{D(x;M) \propto {M\over |x|} K_1(M|x|) \approx \sqrt{\pi/2} 
{M\over |x|^{3/2}} e^{-M|x|} \left(1+ \calo{1\over M|x|}\right)\ .}
The massless propagator is 
\eqn\mprop{D(x,0)\propto {1\over x^2}\ .}
Let $w$ be the perpindicular vector from the origin to the line $xx'$, and
decompose $y$ into components perpindicular or parallel to this line 
as $y=w+z_\perp+z_\parallel$. Since we're working near $xx'$, expand to
leading order in $z_\perp$.  From \exactgf, \feynprop, and \mprop, we find 
\eqn\approxgf{\eqalign{&\int dy \, D(0,y;0) 
D(y,x;M) D(y,x';M) \cr & \roughly\propto 
e^{-M|x-x'|} \int d^4z {1\over |x-y|^{3/2} |x'-y|^{3/2}} \exp\left\{-{M|x-x'|
z_\perp^2 \over 2|x-y_\parallel||y_\parallel-x'| }\right\} {1\over
(w+z)_\perp^2 + z_\parallel^2}\ .}}

Were it not for the last factor, the integral would clearly have a line of
saddlepoints at $z_\perp=0$ along the line $xx'$ governed by small parameter 
$1/M|x-x'|$, just as reasoned above.  However, the last factor in \approxgf\ 
becomes large precisely where the light cone of the source intersects this
line and changes the saddlepoint structure so that there are individual saddlepoints
just off the line in the vicinity of the light cone.  We have not yet completed a full 
treatment of the
resulting (correct) saddlepoint analysis, but in the field theory case we
know, as discussed above, 
that the exact result is zero and any valid saddlepoint analysis
should not contradict this.  The main point of this discussion was
to show that the saddlepoints are not of the form assumed in \BCFM.

\subsec{Wilson loops, reconsidered}

The reasoning of section two could equally be applied in Minkowski space to
argue that measurement of large loops of string allow us to see events at
spacelike separation.  Consider a circular Wilson loop near a point-like
source such that the future light cone of the source passes through the
interior of the loop, intersecting the disk spanning the circle.  The
loop, however, is large enough that it is fully outside the light cone.

\ifig{\Fig\LoopAndCone}{The light cone of the source passes through the
interior of a large Wilson loop, though the loop itself is outside the
light cone.}{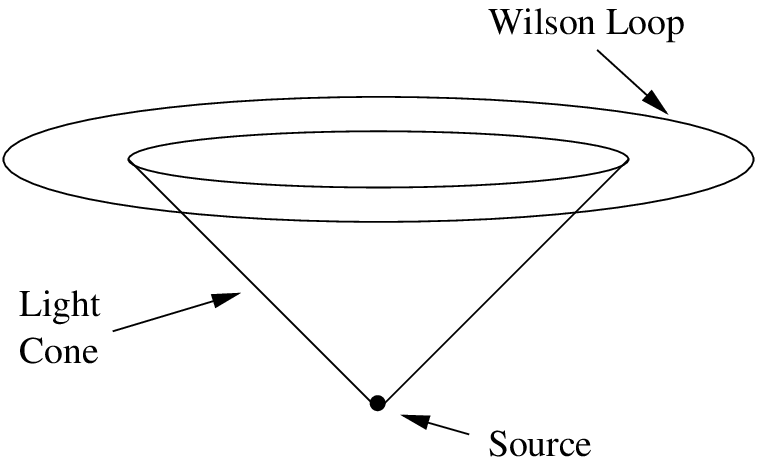}{2.0} 

As emphasized above, to check whether the Wilson loop measures the
effects of the source, we need to compute the off-shell two-point function of
\Impart.  Though we don't know how to do this properly, we expect it
to be represented by an integral over world sheets with
the topology of the disk, with a point-like source vertex operator
$V(0)$ at the origin, with boundary on the curve $x(\sigma)$, and
weighted by the Polyakov action $S_P$: 
\eqn\stringint{\int^{x(\sigma)} \cald X \cald  g \, e^{-S_P[X,g]} V(0)\ .}
Computing this in $AdS_5\times S^5$ is even more problematic, given the
lack of technology for Ramond-Ramond backgrounds.  

Since the exact calculation is difficult, we will try an approximation
in the spirit of \refs{\Mald,\BCFM}.  We assume the integral over all
world sheets can be rewritten, as in \wrongsadd, as an integral over
minimal world sheets, but with the constraint that they are attached at an
{\it arbitrary} 
point $y$ to a dilaton propagator.  The resulting expression is  
\eqn\coneint{\int dy \, e^{-TA(y)} D(0,y)}
where $T$ is the string tension, $A$ is the world sheet area,
and $D$ is the dilaton propagator.

\ifig{\Fig\LoopAndProp}{The integral over all world sheets can be
approximated by integrating over $y$ where the dilaton propagator is
attached to a conical world sheet.}{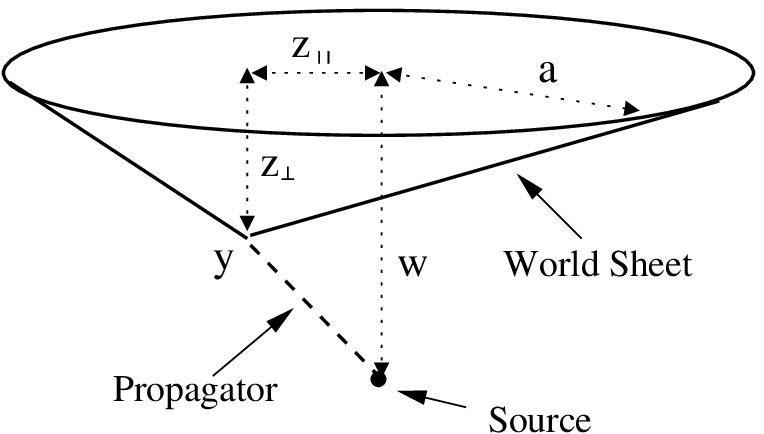}{2.0}

For simplicity, we take the source to lie on the axis of the loop.  For a circular Wilson loop of radius $a$, the minimal world sheet is
a tilted cone whose base is the Wilson loop and whose apex is at
$y$.  As before, let $y = w + z_\perp + z_\parallel$, where $w$ is the
vector from the origin to the center of the loop, $z_\perp$ is the
distance from the plane of the loop, and $z_\parallel$ is the radial
distance from the center of the loop.  The area of the tilted cone is
given by the integral
\eqn\conearea{A = \int^{2\pi}_0 d\theta \frac{a}{2} \sqrt{(a-z_\parallel
\cos{\theta})^2 + z_\perp^2}.}
In the limit where $Ta^2 \gg 1$, the integral \coneint\ appears to be
dominated by world sheets with $z_\perp^2 \ll a^2$.  We can therefore
expand \conearea\ to leading order in $z_\perp^2$ and integrate term by term,
obtaining 
\eqn\approxconearea{A \approx \pi a^2 \left(1 +
\frac{z_\perp^2}{2a\sqrt{a^2-z_\parallel^2}} \right)\ .} 
This reduces \coneint\ to  
\eqn\stringap{e^{-T\pi a^2} \int d^4z \,
e^{-{Tz_\perp^2}/\left(2a\sqrt{a^2-z_\parallel^2}\right)} 
\frac{1}{(w+z)_\perp^2
+ z_\parallel^2}\ .}
Notice the close analogy to the approximate expression for the particle,
\approxgf.

Clearly, as in the case of the particle, there is no longer a
surface of saddlepoints
along the minimal disk spanning the Wilson loop, but
rather there are saddlepoints shifted off this disk near the light cone of the 
source.  Although we have
again not performed a systematic saddlepoint approximation about these, the
strong analogy to the particle case suggests that once correctly computed,
the resulting expression vanishes.

Of course, it would be instructive to attempt to complete a more accurate
calculation of the the correlator of a large Wilson loop -- or loop of
string in bulk language -- with an approximately point-like source.  
It is conceivable that study of the exact off-shell string amplitude
\stringint\ will produce a non-vanishing result.  This faces difficulties,
but may be tractable.  One approach is a careful treatment by an 
intermediate semiclassical approximation working about the correct
saddlepoints, as sketched above.  Another
alternative would be to work directly in AdS space.  In AdS, taking a state
to the boundary in effect corresponds to working on shell, and this
statement may hold equally well for macroscopic string loops.  In this
case, if the initial dilaton can be arranged to be on shell, the issue
might be settled by a completely on-shell calculation in AdS space.  Of
course, a full calculation would require confronting the difficult
problem of Ramond-Ramond backgrounds, but this approach is worth
exploration.  

\newsec{Towards a theory of precursors}

According to the above analysis, calculations to date have not demonstrated
that the large Wilson loops of \SuTo\ serve as precursors, and this reopens the problem of
their correct identification.  We therefore turn to an investigation of
this problem.

\subsec{Field theory locality}

We begin by discussing the problem in the context of a field theory of
a scalar field $\phi$ ({\it e.g.} the dilaton) in AdS.  Let us start
by examining more closely the form of the boundary state $\vert
j\rangle_\partial$ created by the source \source.

First, consider the map between bulk and boundary in more detail.  
Begin with the bulk theory in the supergravity limit, at weak coupling.  
At zeroth order in the coupling, the field $\phi$ has an expansion in terms
of canonically normalized
annihilation/creation operators and mode functions:
\eqn\phiexp{\phi(x) = \sum_{nl\vecm} a_{nl\vecm} \phi_{nl\vecm}({\vecx,\tau})
+ a_{nl\vecm}^\dagger \phi_{nl\vecm}^*({\vecx,\tau})\ .}
Single particle states are of the form
\eqn\singpart{\vert\nlm\rangle_B = a_\nlm^\dagger\vert0\rangle_B \ .}
Likewise, as emphasized in \refs{\BDHM}, 
the corresponding boundary operator should have an expansion of
the form
\eqn\bdyexp{\calo(\ehat,\tau) = \sum_\nlm {\alpha_\nlm\over \sqrt{2\omega_{nl}}}
 Y_{l\vecm}(\ehat)
e^{-i\omega_{nl} \tau} + {\rm h.c.}\ , }
where the unit vector ${\hat e}$ labels a point on the $S^3$ boundary.

In comparing states, the bulk and boundary vacua should correspond,
\eqn\vaccorr{\vac_\partial \leftrightarrow \vac_B\ .}
To get the relation between excited states, use the operator correspondence
\opfield. This becomes
\eqn\crecorr{\alpha_\nlm\leftrightarrow k_{nl} a_\nlm\ }
where $k_{nl}$ are constants given by the asymptotics of the mode functions
$\phi_\nlm$ (see {\it e.g.} the appendix to \refs{\FSS}).
%
%
So the correspondence between single particle states takes the form
\eqn\onestate{ \alpha_\nlm^\dagger \vac_\partial \leftrightarrow 
k_{nl} \vert
\nlm\rangle_B\ .}
Indeed, \crecorr\ relates an arbitrary multi-particle bulk state to a
boundary state.  

We can also read off the relation between the bulk and boundary states
created by an arbitrary source.  This is most easily accomplished by
inverting the relationship \opfield\ to determine the bulk field
corresponding to a given boundary operator.  This is done using the
transfer matrix $M(x,b)$ given in \refs{\BGL, \BDHM}, and we find the relation
\eqn\bulktobdy{ \phi(x) \leftrightarrow \int db \, M(x,b) \calo(b)\ .}
One immediately deduces that the boundary state corresponding to the source 
\bulksource\ is given by the formula \bdystate, with the identification 
\eqn\jtof{f(b)=\int dV_x \, j(x) M(x,b)\ . }

Now, the question is what kinds of boundary operators with support only outside
the light cone of the source can detect the state $\vert j\rangle_\partial$.  
Clearly
neither $\calo$ nor any of its derivatives do, since $\calo$ corresponds to
the bulk field through \opfield, and the bulk fields commute outside the
light cone.  Within the context of this simple field
theory model, the only way to get operators that play the role of
precursors is to identify other local operators on the boundary that are
sensitive to the data in the state $\vert j\rangle_\partial$.

An example that has the appearance of a cheat is if there is another set of
local observables that can be written in terms of the $\alpha_\nlm$'s in
the form
\eqn\newfield{\calo'(\ehat, \tau) = \sum_\nlm c_\nlm {\alpha_\nlm\over
\sqrt{2\omega_{nl}}}
 Y_{l\vecm}(\ehat)
e^{-i\omega_{nl} \tau} + {\rm h.c.}\ .}
For generically chosen coefficients $c_\nlm$ these operators will not
commute with the field operators outside the light cone.  The reason this
looks like a cheat is that the relationship between $\calo$ and $\calo'$ is
of course highly nonlocal.

It should be recalled, however, that in the context of AdS/CFT, operators
like $\calo$ are composites of the fundamental Yang-Mills boundary
fields. This leads us to the question of whether there are other
gauge-invariant operator combinations of these fields that are able to
measure the state \bdystate\ outside the light cone.

A toy model for such a possibility was given in \refs{\PST}.  Polchinski,
Susskind, and Toumbas considered modelling the boundary theory as a theory
of 
$N\times N$ matrix scalar fields $\psi_{mn}$.  In terms of these fields, the
boundary state \bdystate\ can be thought of as a squeezed state.  Indeed,
for free scalar fields, an obvious analog to ${\rm Tr} \, F^2$ of Yang-Mills is an
operator of the form
\eqn\toyop{\calo = {1\over N}{\rm Tr}\left[(\nabla\psi)^2\right]\ .}
The state \bdystate\ then takes the form of a squeezed state in terms of
the annihilation and creation operators
\eqn\bannc{b_{mn}(k)\ ,\ b_{mn}^\dagger(k)}
of for the $\psi_{mn}$ fields:
\eqn\tbound{ \vert j\rangle_\partial = \exp\left\{ {1\over 2}\int d^3k \, d^3k' F(k,k') b^\dagger_{mn}(k)  b_{nm}^\dagger(k') 
\right\} \vac_\partial\ }
as in \PST.  

The authors of \PST\ investigate the problem of detecting such a state using a bilocal, bilinear in the fields $\psi_{mn}$ and suggest that this is possible.  In our language, one should investigate expressions of the form
\eqn\commtest{\left[ \psi_{mn}(b) \psi_{nm}(b'), \int db'' M(0,b'') \calo(b'')\right] }
(or analogous expressions with the source coupled to the stress tensor) for $b, b'$ outside the light cone.\foot{It would also be interesting, though not convincing because of gauge non-invariance, to exhibit a non-vanishing commutator between the source and a single  $\psi_{mn}$.}

The bilinear in \commtest\ is the analog of a certain kind of decorated Wilson loop in the limit of zero coupling, and a non-vanishing result for \commtest\ would be a potentially interesting indication that such decorated loops play a role as the precursors.  This possibility is under investigation.  Another interesting question is to better understand the relationship of such decorated loops to the AdS boundary S-matrix\refs{\BGL, \BSM}.  Assuming these loops corresponding to elements of the boundary S-matrix, they shouldn't exhibit any bulk acausality that should not be evident in that S-matrix.  Certainly, for generic
low-energy scattering experiments in the string theory of the bulk, we do
not expect to be able to explicitly exhibit this nonlocality. 

In the next subsection we turn to a discussion of physical situations where we expect that
nonlocality {\it should} be manifest.  

\subsec{Saturation a string/gravity locality bound}

Another place to look for clues regarding the precursors is to return to
the motivations for holography.  It is believed that, in contexts where
strong gravitational effects are relevant, the number of fundamental
degrees of freedom are drastically reduced in a fashion conflicting with
na\"\i ve locality.  One situation where this is thought to occur
is black hole formation.
Therefore, in searching for origins of the
nonlocal precursors, we should consider situations where locality breaks 
down due to black hole formation.

A likely connected statement (through black hole/string 
correspondence\refs{\HoPo}) is the belief that
when string effects are important, na\"\i ve locality is again
violated, as has been seen for example in string modifications of the 
uncertainty principle\refs{\GrMe}.

In order to understand in what situations holographic bounds begin to
affect causality, recall that in field theory causality is formulated as
the statement that fields commute at spacelike separations:
\eqn\classcomm{ [\phi(x),\phi(x')] =0\ ,\ (x-x')^2>0\ .}
However, we expect that the corresponding equation in string theory, schematically
\eqn\Strcomm{\left[\Phi[x(\sigma)],\Phi[x'(\sigma)] \right] =0}
does not hold when gravitational or string
effects become strong.  Consider the limit where the curves $x(\sigma)$ and
$x'(\sigma)$ are nearly point-like; we expect commutativity to fail
when we consider modes of the operators that
are sufficiently high energy to create a string or black hole (or other
M-theoretic object) occluding the
points $x$ and $x'$.  Of course, the operators $\Phi[x(\sigma)]$ include all
possible momenta, but to apply this criterion we can work in a wavepacket
basis\refs{\Hawk,\GiNe} in which states have nearly definite
momenta and positions
satisfying the Heisenberg uncertainty relation $\Delta x\Delta p\roughly>1$.

These ideas lead us to the following 

{\it Criterion for a locality bound}:  

\noindent Consider two particles (or strings) of momenta $p_1$ and $p_2$
colliding with impact parameter (measured in the center of mass frame) $b$.
These will be
said to saturate the locality bound if the collision is
sufficiently energetic to create either a 
string or black hole with size larger than
$b$.  

A rough condition for this is that the energy simply be large enough to
form a black hole or string larger than $b$ -- which effect is most
important depends on the string coupling.  So in spacetime dimension $D$ 
this condition becomes\foot{With appropriate modifications 
in case of creation of other fundamental
extended objects.}
\eqn\cmholo{E_{cm}> {\rm min} \left(b/l_{st}^2, b^{D-3}/g_s^2
l_{st}^{D-2}\right) \
,} 
where 
$l_{st}$ is the string length.  Of course interaction, form factor,
etc. effects are expected to modify this bound, particularly at large
energy/impact parameter; we might expect the correct bound from string
production to be somewhere between the two values in \cmholo.

Note that the spacetime uncertainty relation\refs{\Yone} follows as a
consequence of our estimate \cmholo\ and the statement that 
a process confined to a time
interval $\Delta t$ must have energy $E\roughly> 1/\Delta t$. Combining
these implies 
\eqn\stringunc{\Delta t > {\rm max} \left( l_{st}^2/ \Delta x, g_s^2
l_{st}^8/ (\Delta x)^7\right)}
for $D=10$, as in \Yone.  Note also that, as in \Yone, at least according
to these estimates the crossover between string dominance and black hole
dominance occurs at scales
\eqn\bscale{b\sim g_s^{1/3} l_{st}\ ,}
the Planck length of eleven dimensional M-theory.


For simplicity \cmholo\ has been given in terms of flat space kinematics,
but the same basic physical 
principle should determine where locality bounds are
saturated in AdS, and indeed in the limit of large AdS radius $R$ the
statements should correspond.  Because of the complications of AdS
kinematics, let us investigate the bound in the simple picture in which
large-radius AdS
is represented as a cavity of radius $R$, with a flat internal metric.  In
this situation it is straightforward to get a feeling for which
configurations saturate our bound.

Indeed, consider a lightlike particle with rectangular momentum
$(E_1,E_1,{\vec 0})$ emitted from the center of the cavity at time $T=0$.  
Suppose a second particle is travelling in the opposite direction,
with momentum $(E_2,-E_2,{\vec 0})$, and is located in the vicinity of the
boundary at $R$, with separation transverse to the momenta,
also at time $T=0$.  Let us ask what energy $E_2$ is
required to saturate the estimate \cmholo.  In this case the CM impact
parameter is $\approx R$, and the CM energy is $2\sqrt{E_1E_2}$, 
so our estimate
states
\eqn\exbound{ E_2\roughly> {1\over E_1} {\rm min} \left({R^2\over
l_{st}^4},  {R^{14}\over g_s^4 l_{st}^{16}}\right)}
(for black holes of radius $\roughly<R$, we take $D=10$).
Similar statements can be readily derived for other configurations.  Note
that for large $R$ and small $g_s$, this suggests that the relevant bound
is from string creation.

In short, 
while it is not clear {\it how} in detail 
the information is holographically 
encoded and exhibited, the above physical criterion serves as a guide to
when locality should fail and
holographic effects are expected to become important.  If there is indeed an
underlying unitary and holographic theory (such as M-theory), this
criterion indicates where it should cease to appear local and start to
appear holographic.  The estimate
\cmholo\ clearly neglects important effects, but gives a rough idea of
the nature of such a locality bound.  

Turning to the boundary theory, we can now use the correspondence between
the  AdS boundary S-matrix and the boundary correlators to infer which
correlators in the boundary theory we expect to exhibit effects that
violate na\"\i ve bulk locality.  As discussed earlier, the Wilson loops,
or equivalently, via \Azam, the set of all local operators, form a basis for
the boundary observables, but the question is what combinations of these
operators are most sensitive to effects that begin to saturate our
locality bound.

From the above discussion, we expect these to be projections onto operators
that correspond to creation of large, high-energy intermediate states, for
example black holes or large strings.
Two obvious
possibilities exist. One is to consider the high-energy components of local
operators $\calo(x)$ (or equivalently the high-energy components of Wilson
loops).  This corresponds to resolving variations of boundary correlators
on very short time/distance scales.  Alternatively, one might consider
correlators with a very large number of softer operators that combine to
give a large energy.

It is not clear that such nonlocalities would be manifest at string tree
level.  The authors of \refs{\LPSTU} attempted to exhibit such effects 
in a
three-point string tree-level calculation but could not conclude that 
what
they saw wasn't a gauge artifact.  These effects may require
higher loop or non-perturbative calculations, which would certainly make
sense if intermediate black holes or large strings play a role.

It is also not clear that a large Wilson loop is sufficient to probe these
nonlocalities.  A Wilson loop is not intrinsically high energy any more
than the field operator $\phi(x)$ is in field theory; rather, it involves a sum
over all energies.  One expects to be sensitive to nonlocalities by
projecting onto certain high-energy components of these operators.

Of course, concrete calculations that exhibit such nonlocal results,
particularly from very high energy operators, or large collections of soft
operators, may well be rather difficult if indeed the nonlocality results
from higher-loop/non-perturbative effects.  We leave this problem for the
future.

\newsec{Black hole information}

In light of the above, we now revisit the original motivation of using the
precursor fields to ``see'' inside a black hole.

\subsec{Large Wilson loops and flossing black holes}

First, consider the possibility that 
an improved version of the tree-level calculation of \SuTo\ indeed reveals
acausal effects; then we should obviously consider applying it to black
holes.  Consider, for example, a black hole of radius $r_h\ll R$ sitting at
the center of AdS, and suppose that we wish to measure whether a bomb
dropped into the black hole has detonated or not.  According to \SuTo, 
we could hope to do so
by measuring a large Wilson loop at the boundary.  By the criterion of
section three, 
this Wilson loop would be able to measure a source inside
the black hole if its spanning minimal surface crosses the horizon and
intersects the future light cone of the source.
Clearly, a Wilson loop that is a great circle on the $S^3$ boundary of AdS
will, by symmetry, have a spanning surface that cuts through the center of
the black hole.  If we move this circle off the equator of the sphere, then
eventually it will not enter the horizon; we expect this to happen when the
circle reaches a latitude of order 
\eqn\longit{\Delta\theta \sim r_h/R\ .}  
Although
our preceding analysis demonstrates that this surface is not the correct
saddlepoint for the functional integral, the correct saddlepoint is
a deformation of this surface.  {\it If} this saddlepoint yields a
non-vanishing result for such measurements, then we'd expect that to occur
for Wilson loops in the range \longit\ about great circles.

\ifig{\Fig\FlossSpacelike}{Measuring an instantaneous Wilson loop
corresponds to a virtual world sheet which goes through the black
hole.}{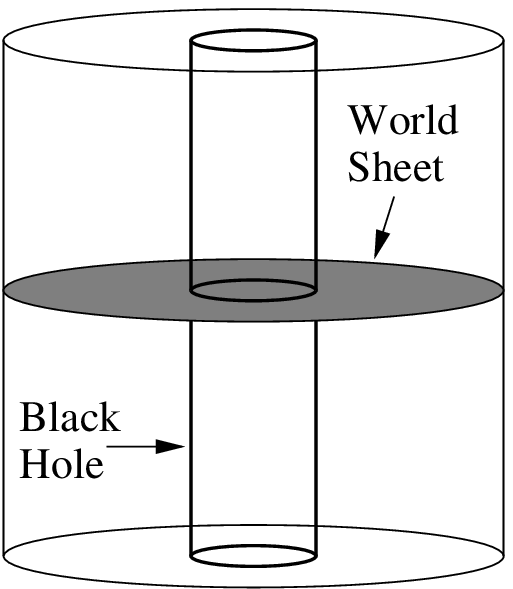}{2.0}

However, in addition to the preceding arguments, 
there are physical reasons to be suspicious of such claims.
Consider the picture of a world sheet instantaneously slicing a black hole,
as in fig.~7.  This process involves a virtual string,
but is dual to another (idealized) process involving a real string state.
This is a process in which an observer near the boundary 
creates a piece of string, then
stretches it to macroscopic scales, slices it through the black hole, and
then shrinks it back down at the opposite side of AdS, as shown in
fig.~8 . 

\ifig{\Fig\FlossTimelike}{A timelike world sheet corresponds to a string flossing the black hole.}
{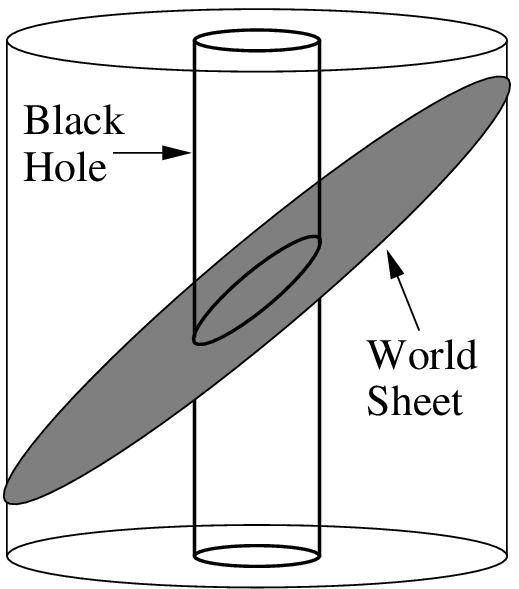}{2.0} 

So, an obvious question is whether one expects to be able to mine information
from a black hole by this process of flossing it with a string.  If the
answer is negative, it seems even more unlikely that the information is
manifested in the far off-shell version of this process.

We are skeptical that such an effect can be seen in a tree level
calculation.  Indeed, as discussed in \refs{\LaMa}, which considers a
configuration with a stationary string threading a black hole, the string
is expected to inherit the causal structure of the spacetime.  In this
situation, the only string excitations that make it out to an observer at
infinity are the Hawking radiation of oscillation modes on the string, and
these do not contain information about the state inside the horizon or any
perturbations of it by classical sources inside  the black hole.  Trying to
pull the string back out of the black hole adds another layer of
difficulty;  this should not be possible without the string
breaking off a closed loop that remains inside the black hole.\foot{The
string trajectory that corresponds to pulling the string completely back
out of the black hole without leaving a loop 
clearly can't satisfy the classical string
equations of motion, and we'd expect the neighboring
trajectories to contribute rapidly varying and cancelling phases to the
corresponding quantum mechanical amplitude.}  We'd expect
this closed loop to contain any information from inside the black hole and
the remaining external string state to be insensitive to the internal state
of the black hole, at least at string tree level.

Again, at higher level in $g_s$ one certainly might imagine seeing interesting 
effects, if the basic ideas of holography are correct and realized through
stringy corrections.  If so, it is plausible that Wilson loops in the
range \longit\ are indeed sensitive to those effects, although
we will advocate an alternative viewpoint.

\subsec{Black holes and holography}

In parallel to the discussion of the preceding section,
we could ask in greater generality where in the
boundary theory we might expect to see the information contained in the
interior of a black hole.  
To address this, we recall two facts.  The first is the AdS/CFT
correspondence between the AdS boundary S-matrix and the CFT
correlators, outlined above.  
Secondly, for a large black hole in a much larger AdS space, the bulk dynamics
should be closely approximated by a black hole in flat space.  We expect
intermediate states with large black holes to arise in specific blocks of
the S-matrix.  One example is a matrix element with
sufficient energy in the initial state,
focussed into a region of order its Schwarzchild radius; in this case, the
final state is expected to include a very large number of outgoing soft
quanta -- the Hawking radiation -- and if the information is contained
therein, it should be in subtle correlations between this large number of
quanta.  This number should be 
$\calo(A)$ for an intermediate black hole of area $A$.

Mapping these statements to the boundary theory, we might
investigate black holes through correlators that have operators
corresponding to energetic
and narrowly focussed incoming states, with center of mass energy $E$, 
and a large number ($\calo(E^{8/7})$ in ten dimensions) 
of operators corresponding to the soft outgoing quanta.  The subtle
relative phase information in these would describe the black hole
information, which may be difficult to see otherwise.  It could also be, in
line with our earlier arguments, that certain other correlators with few,
but very high energy operators, arranged so that they start to saturated
locality bounds, as in the preceding section, 
would be sensitive to this information.

Unfortunately, with the present state of our knowledge this proposal does
not give further details about how to escape the black hole information
paradox; in a sense it is simply mapping our earlier attempts at a
holographic explanation of its resolution into the AdS/CFT arena.

\newsec{Conclusion}

The question of identifying the precursor variables in AdS/CFT is an
important one, both because of its relevance to understanding the detailed
relation between approximately local bulk physics and boundary physics and
because of its promise to finally explain how holography resolves the black
hole information paradox.  In this paper we have investigated the proposal
of \SuTo\ that large Wilson loops are the precursors and found a difficulty;
specifically, the analysis of \BCFM\ that was used misidentified the
saddlepoint dominating the functional integral over world sheets.  We gave
an alternative proposal, in which the precursors are related to
observables that saturate a certain locality bound.  The physical idea
underlying this bound is that one should not be able to make observations
that involve concentrating an amount of energy within a region smaller
than the corresponding Schwarzschild radius or in a region small
enough such that other large objects, such as strings, with size
comparable to the region, will be created.  This bound can be saturated either
by individual high-energy operators or by collections of soft operators
with large total energy.  This suggests to us where to look in order to
understand how AdS/CFT resolves the black hole information paradox, but
unfortunately does not yet tell us how to make detailed calculations
exhibiting the solution.

\bigskip\bigskip\centerline{{\bf Acknowledgments}}\nobreak

The authors wish to thank D. Gross, G. Horowitz, V. Hubeny, A. Lawrence,
J. Polchinski, J. Preskill, and L. Susskind for valuable conversations. 
Parts of this work were carried out at Caltech,
and the Caltech/USC Center for Theoretical Physics,
whose support and hospitality are gratefully acknowledged.  This work was
supported in part by the Department of Energy under
Contract DE-FG-03-91ER40618, and by the National Science Foundation under
Grant No. PHY99-07949.

\listrefs
\end